# Electron-plasmon interaction induced plasmonic-polaron band replication in epitaxial perovskite SrIrO$_3$ films

Zhengtai Liu [a,b], Wanling Liu [a,b,c], Ruixiang Zhou [c], Songhua Cai [d,e], Yekai Song [a,b,c], Qi Yao [c], Xiangle Lu [a,b], Jishan Liu [a,b], Zhonghao Liu [a,b], Zhen Wang [e,f], Yi Zheng [e,f], Peng Wang [d,e], Zhi Liu [a,c], Gang Li [c,\*], Dawei Shen [a,b,\*]

[a] State Key Laboratory of Functional Materials for Informatics, Shanghai Institute of Microsystem and Information Technology (SIMIT), Chinese Academy of Sciences, Shanghai 200050, China
[b] Center of Materials Science and Optoelectronics Engineering, University of Chinese Academy of Sciences, Beijing 100049, China
[c] Division of Photon Science and Condensed Matter Physics, School of Physical Science and Technology, ShanghaiTech University, Shanghai 200031, China
[d] Laboratory of Solid State Microstructures, College of Engineering and Applied Sciences, Nanjing University, Nanjing 210093, China
[e] Collaborative Innovation Center of Advanced Microstructures, Nanjing University, Nanjing 210093, China
[f] Department of Physics, Zhejiang University, Hangzhou 310027, China

\*Corresponding authors
Email addresses: ligang@shanghaitech.edu.cn (Gang Li), dwshen@mail.sim.ac.cn (Dawei Shen)

---

**Abstract**
Electron-boson interaction is fundamental to a thorough understanding of various exotic properties emerging in many-body physics. In photoemission spectroscopy, photoelectron emission due to photon absorption would trigger diverse collective excitations in solids, including the emergence of phonons, magnons, electron-hole pairs, and plasmons, which naturally provides a reliable pathway to study electron-boson couplings. While fingerprints of electron-phonon/-magnon interactions in this state-of-the-art technique have been well investigated, much less is known about electron-plasmon coupling, and direct observation of the band renormalization solely due to electron-plasmon interactions is extremely challenging. Here by utilizing integrated oxide molecular-beam epitaxy and angle-resolved photoemission spectroscopy, we discover the long sought-after pure electron-plasmon coupling-induced low-lying plasmonic-polaron replica bands in epitaxial semimetallic SrIrO$_3$ films, in which the characteristic low carrier concentration and narrow bandwidth combine to provide a unique platform where the electron-plasmon interaction can be investigated kinematically in photoemission spectroscopy. This finding enriches the forms of electron band normalization on collective modes in solids and demonstrates that, to obtain a complete understanding of the quasiparticle dynamics in 5*d* electron systems, the electron-plasmon interaction should be considered on equal footing with the acknowledged electron-electron interaction and spin-orbit coupling.

**1. Introduction**

Determining how electrons behave when they interact in crystals is a principal problem in condensed matter physics. To solve this complex many-body problem, the quasiparticle picture is most often applied to describe the electronic structure of solids [1]. In this approximation, the strongly correlated electrons are considered weakly interacting quasi-electrons/-holes, which are composed of bare particles, the screening cloud of electron-hole (e-h) pairs, and collective excitations. Owing to their analogous independence, an energy and lifetime renormalization resulting from the interaction between electrons and collective excitations must be introduced to the bare band.

In photoemission spectroscopy, the width of the spectral peak from photoelectron energy distribution curve (EDC) is inversely proportional to the quasiparticle lifetime, which makes this technique a straightforward probe of various couplings of electrons with collective excitations [2]. The characteristics of electron-phonon/-magnon interactions in photoemission spectra have been thoroughly investigated [3–12], while less is known about the interaction between electrons and plasmons, which are typical collective charge density oscillations of the electron gas resulting from the long-range nature of the Coulomb interaction. In the 1960s, Lundgvist theoretically predicted a new composite particle [13], a plasmaron, caused by the strong coupling between holes and plasmons, which would give rise to distinct satellite band structure from the conventional quasiparticle feature in photoemission spectra. Furthermore, with the growing interest in plasmonic devices [14–16], more recent proposals have suggested that electron-plasmon coupling should lead to plasmonic-polaron band-structure replica, which is analogous to the polarons in the conventional theory of electron-phonon interactions, manifesting as broadened copies of the valence bands shifted by the plasmon energy [17–20]. However, photoemission spectroscopic probing of conventional plasmaron or plasmonic-polaron replica located at high binding energy (~10 eV) may be hindered by the relatively poor momentum resolution and possible mixup of weak plasmonic satellites and strong quasiparticle peaks [18]. To date, plasmonic satellites have only been reported in the integrated photoemission spectra of core/valence bands of Na [21] and Si [22] and have been partially observed in a narrow region around the Dirac point of graphene [23, 24]. Comprehensive dispersion information of pure electron-plasmon coupling-induced replica bands near the Fermi level ($E_F$) is lacking. Such information could provide straightforward insights into the complete picture of the quasiparticle dynamics.

Here, we employ an integrated system of oxide molecular-beam epitaxy (MBE) and angle-resolved photoemission spectroscopy (ARPES) to present the long sought-after plasmonic-polaron replica of bands of epitaxial perovskite SrIrO$_3$ in the vicinity of $E_F$, analogous to band replication resulting from Fröhlich interaction between electrons and longitudinal-optical (LO) phonons in polar materials [8–11, 25, 26]. Further electronic structure evolution studies that consider both dimensionality and carrier concentrations, together with sophisticated *ab initio* calculations, could confirm the pure and intrinsic electron-plasmon coupling origin of these replica bands.

## 2. METHODS

*2.1. Growth and characterizations of thin films*

Perovskite-type SrIrO$_3$ thin films were deposited on (001) SrTiO$_3$ substrates at a substrate temperature of 560°C with a background pressure of $2\times10^{-6}$ Torr of distilled ozone. The growth temperature was verified by optical pyrometry. Strontium and iridium were evaporated from an effusion cell and an electron beam evaporator, respectively, and the co-deposition method was applied where both elements' shutters were open for the duration of the growth. The strontium and iridium fluxes were approximately $1.2\times10^{13}$ atoms/(cm$^2$·s). Fluxes were checked using a quartz crystal microbalance before and after the deposition. During the growth process, the *in situ* reflection high-energy electron diffraction (RHEED) intensity oscillations and patterns were collected to monitor the overall growth rate and surface structure. The crystal structure of thin films was examined by X-ray diffraction (XRD) using a high-resolution Bruker D8 discover diffractometer. In addition, a cross-sectional transmission electron microscope (TEM) sample was prepared by focused ion beam. Atomic resolution scanning transmission electron microscope (STEM) observation was carried out on a double aberration-corrected Thermo Fisher Titan G2 60-300 S/TEM with accelerating voltage of 300 kV. The probe-forming semi-angle was 25 mrad. A high-angle annular dark field (HAADF) detector with a collection semi-angle ranging from 79.5 to 200 mrad was used.

*2.2. Angle-resolved photoemission spectroscopy*

Thin films were transferred to the combined ARPES chamber through an ultrahigh vacuum buffer chamber (~$2\times10^{-10}$ Torr) for measurement immediately after growth. This ARPES system is equipped with a VG-Scienta R8000 electron analyzer and a helium discharge lamp. Data were collected at 12 K under ultrahigh vacuum conditions ($8\times10^{-11}$ Torr). The angular resolution was 0.3°, and the overall energy resolution was set to 10 meV. During the measurement, films were stable and did not show any sign of degradation.

*2.3. Measurement of the transport properties*

The measurement of thin film electrical transport properties was performed in a physical property measurement system (Quantum Design, Dynacool) with a standard six-terminal method. The temperature dependence of resistivity was measured at a constant current of 10 μA from 2 to 300 K. The Hall and magnetoresistance (MR) measurements were taken with a magnetic field up to 9 T applied perpendicular to the films surface at 10 K. To separate the Hall contribution from that of longitudinal MR, an antisymmetrization procedure was performed by separating the positive and negative field sweep branches.

*2.4. First-principles calculations*

Our theoretical results and analysis were based on different levels of first-principle calculations. To clarify the effect of lattice vibration, a fully relaxed crystal structure and the phonon spectrum were determined using the Quantum ESPRESSO package [27] with a kinetic energy cutoff of 680 eV and 4×4×4 momentum sampling in the first Brillouin zone (BZ). Based on that, the phonon-mediated electronic band structure was calculated using the electron–phonon interaction through Wannier functions (EPW) program [28] with the Wannier90 [29] package. The interaction effect on the electronic structure was studied by combining WIEN2k [30] and dynamical mean-field theory (DMFT) [31]. Theoretical calculations are described in detail in the Supplementary materials.

## 3. Results and discussion

Since bulk $SrIrO_3$ prefers to crystallize in the 6*H*-hexagonal structure rather than the perovskite phase [32, 33], we performed *in situ* ARPES measurement on epitaxial $SrIrO_3$ films prepared by reactive MBE to investigate the detailed quasiparticle dynamics in this perovskite iridate. In Fig. 1a, the insets show the RHEED patterns of the $SrTiO_3$ substrate and 20 unit-cell (uc) $SrIrO_3$ films, respectively. The RHEED image of $SrIrO_3$ films exhibits a sharp streak and Kikuchi lines, suggesting high surface perfection. In addition, as shown in Fig. 1a, uniform periods of RHEED oscillation curve throughout the entire deposition duration demonstrate the persistent layer-by-layer growth of thin films. Moreover, as shown in Fig. 1b, the cross-section HAADF STEM image indicates the absence of domains and perfect epitaxial growth of $SrIrO_3$ films across the interface. According to the close-up shown in the right panel of Fig. 1b, all atoms are atomically aligned as expected in films. Together with detailed XRD measurements (see the Supplementary materials for details), the above characterizations demonstrate the perfect surface and crystal structure of our films, which can guarantee their intrinsic properties to be revealed by further *in situ* ARPES.

We examined band dispersions of $SrIrO_3$ films along high-symmetry directions with 21.2 eV photons, which have been reported to correspond to the $k_z$ plane close to *Z–R–U* in the BZ [34, 35]. As shown in Fig. 1c, three hole-like bands *α*, *β*, and *γ* barely intersect $E_F$, which is consistent with previous experiments [34, 35]. For bands *α* and *γ*, we note that their bandwidths are only around 170 meV and 190 meV, respectively. Note that band *β* disperses by 290 meV as well. The extremely narrow bandwidth of these bands has generated significant research interest because it suggests unconventional electronic structure evolution upon dimensionality in Ruddlesden-Popper series of iridates [34, 35]. We will discuss these narrow bands in detail later in this section.

The most unexpected discovery in perovskite SrIrO$_3$ photoemission data is the presence of replica bands, labeled $α'$ and $γ'$, as shown in the photoemission intensity plot (Fig. 1c) and more clearly in the corresponding second derivative in energy (Fig. 1e). Intriguingly, except for a common redshift of binding energy by ~230 meV, all replica bands fully duplicate the features of the main bands, including band dispersions and terminations. We note that the main bands of $α$ and $γ$ can be well reproduced in our first-principles calculations, in which no collective excitations have been included. In these calculations, rather than any bulk or surface states, an energy gap of approximately 300 meV is located at the binding energy of those observed replicas (see the Supplementary materials for details). We can even distinguish a higher order replica $α''$ located at a binding energy of approximately 460 meV below the main band from the second derivative in energy, as shown in Fig. 1e (see the Supplementary materials for details). In addition, along other high-symmetry directions (*Z-U* and *U-R*), as shown in Figs. 1d and Fig. S7 (online), photoemission data also demonstrate similar band duplication behavior for bands $α$ and $γ$, with nearly the same energy shift. Figure 1g and h show corresponding EDCs along these high-symmetry directions, which further confirms the presence of such band replication.

One plausible explanation for replica bands is the presence of subbands due to quantum confinements occurring in thin films. However, the equal energy spacing of different order replica bands is inconsistent with the typical behavior of subbands originating from quantum confinement [36, 37]. To completely rule out the possibility of quantum well states, we have performed a careful comparison of the low-lying electronic structure of epitaxial SrIrO$_3$ films with different thickness. As illustrated in Fig. 2a, the thickness of our films can be well monitored and then precisely controlled in a layer-by-layer manner. Both photoemission intensity plots (Figs. 2b–e) and corresponding second derivatives in energy (Figs. 2f–i) indicate that the energy shift between the main and replica bands remains unchanged for films ranging from 10 to 35 uc thick, as summarized in Figs. 2j and 2k. This negligible thickness dependence of the energy shift, together with the fact that all replica bands terminate at nearly the same momenta as the main bands, is in sharp contrast with typical quantum well states in transition metal oxides [38]. Another possible explanation for these weak intensity bands is the modulation effect of Ir-O octahedral distortions on the low-lying electronic structure of SrIrO$_3$ films [35]. In fact, although in our first-principles band-structure calculations we have run a complete relaxation procedure of lattice constants and fully considered the Ir-O octahedral rotations, we have not observed any signature band replication at a binding energy of approximately 250 meV (see the Supplementary materials for details). We carefully checked previous ARPES work [35] and found a pronounced band gap along the *R–U* direction appearing at a similar binding energy in their first-principles calculations. In addition, the HAADF STEM image shown in Fig. 1b confirms the homogeneity of our films, which could further exclude the possibility that such replicas would be introduced by the inhomogeneity of some domain effects previously reported in thin films [39].

After excluding external origins of band replication, we may attribute such replica band behavior to distinct bosonic shake-off excitations in the case of strong electron-boson coupling [40, 41]. For relatively localized electrons with narrow band width, when such electrons are dressed with bosons in a long-range form, the interaction between electrons and bosons would lead to the emergence of satellites in photoemission spectra. Our finding is rather reminiscent of the two-dimensional Fröhlich polarons observed in oxides, e.g., perovskite $SrTiO_3$ and anatase $TiO_2$ [10, 11], the FeSe/$SrTiO_3$ interface [26], and Graphene/h-BN van der Waals heterostructures [8], in which narrow bands around $E_F$, along with phonons with larger energy than their band width, result in distinguishable shake-off effects.

Intriguingly, we found that the duplication of low-lying quasiparticle bands of perovskite $SrIrO_3$ is sensitive to the carrier concentration, which is in striking contrast to the common behavior of a Fröhlich polaron. Through gradual lanthanum substitution of strontium by oxide MBE, as schematically illustrated in Fig. 3a, we can effectively tune the carrier density $n$ of epitaxial $La_xSr_{1-x}IrO_3$ films, which can then be determined by Hall measurements (see the Supplementary materials for details). Thus, the evolution of replica bands with carrier concentration $n$ could be obtained, as illustrated by both photoemission intensity plots (Figs. 3b–d) and corresponding second derivatives (Figs. 3e–g). We found that the energy of replica bands decreases with $n$ increasing, which can be more straightforwardly verified by a direct comparison of integrated EDCs around $R$, as shown in Fig. 3h. Here, the background of an individual EDC has been subtracted, and the peaks of the main bands have been aligned for better comparison (see the Supplementary materials for details).

Quantitatively, replicas are separated from main bands by 230 meV for as-grown $SrIrO_3$ films, whereas for films with $n = 1.06 \times 10^{20}$ cm$^{-3}$, we obtain an energy difference as large as 270 meV, as shown in Figs. 3d and g. For previously reported Fröhlich polarons, the replica feature is believed to be caused by the forward electron-phonon scattering, and the band-replica separation should be approximately equal to the phonon frequency [25, 42]. However, our light substitution of strontium is not expected to alter the phonon frequency in $La_xSr_{1-x}IrO_3$ significantly. In addition, a band-replica separation larger than 200 meV is far beyond the highest $E_u$ LO phonon in transition metal oxides (~100 meV) [43]. In fact, we have performed a comprehensive electron-phonon interaction calculation for epitaxial $SrIrO_3$ films. On one hand, the calculated phonon spectrum along high-symmetry directions indicates that no phonons higher than 90 meV would exist in this compound (Fig. 4a). On the other hand, as shown in Fig. 4b, the calculated $k$-resolved spectral function combined with electron-phonon coupling could not reproduce such band replication. Thus, the conventional Fröhlich picture based on strong electron-phonon coupling cannot account for the shake-off band structure discovered in perovskite $SrIrO_3$. Moreover, we have also performed state-of-the-art many-body calculations based on the dynamical mean field theory with electron correlations and have not observed any signature of a replica band at $R$-point between −0.4 to −0.1 eV, which confirms that the replica band $\gamma'$ is not of a correlated origin (see the Supplementary materials for details). Alternatively, a recent theoretical work [40] has suggested that the strong coupling with plasmons may also lead to the similar distinctive replica structures in the angle-resolved spectral function of semiconductors with sufficiently high dopant concentrations, which is reminiscent of the plasmonic-polaron bands observed in valence bands of semiconductors at much higher binding energy [22, 44]. A recent experimental study [45] has also discovered the polaronic state in a doped EuO semiconductor. However, the crossover from electron-phonon to electron-plasmon coupling is a barrier for us to study the pure plasmon-polaron. In this regard, the prominent change of band-replica separation energy observed in (Sr,La)IrO$_3$ indeed results from the interaction between the relatively localized electrons [35] near $E_F$ and collective oscillations of free electron gas density with high tunability of frequencies via carrier doping.

In the three-dimensional case, ordinary plasmons follow the dispersion relationship

$$\omega(q) = \omega_{\mathrm{p}} + \frac{3v_{\mathrm{F}}^2 q^2}{10\omega_{\mathrm{p}}}, \qquad (1)$$

and

$$\omega_{\mathrm{p}} = \sqrt{\frac{ne^2}{m\varepsilon}} \qquad (2)$$

where $q$, $e$, $m$, $\varepsilon$, and $v_F$ are the plasmon momentum, electron charge, effective carrier mass, dielectric constant, and Fermi velocity of solids, respectively [46] (see the Supplementary materials for details). Notably, the plasmon energy disperses rather weakly with the momentum, which, remarkably, is analogous to optical phonons. For ordinary metals, typically, the plasmon frequency is rather high (~10 eV). However, for as-grown semimetallic SrIrO$_3$, $n$ is only 8.5×10$^{19}$ cm$^{-3}$, which complies with our expected theoretical calculation results (see the Supplementary materials for details) and approximately three orders smaller than those of ordinary metals, where the corresponding plasmon energy could be evaluated to be approximately 260 meV. This energy is markedly consistent with that of the band-replica separation we measured. Furthermore, the evolution of the energy spacing of replica bands upon $n$ qualitatively matches the change of plasmon mode frequencies deduced through Eqs. (1) and (2) (Fig. 3i), suggesting that our finding is indeed the long sought-after plasmonic-polaron replica of conduction bands in the vicinity of $E_F$.

In the formation of shake-off in photoemission spectra, an important prior condition is that the electron-boson interaction should be dominated by the coupling of electrons with long-wavelength bosons [20, 25]. In the monolayer FeSe film, this prerequisite is satisfied by the significant screening induced by the interface, and the coupling between the FeSe electrons and the SrTiO$_3$ phonon is sharply peaked at zero momentum transfer [26]. With Fröhlich polaronic materials, the electron-phonon coupling matrix elements exhibit a 1/|**q**| singularity, which is rooted in the electric field generated by the finite Born effective charges of the ions; thus, the interaction would be strongly peaked in the forward scattering direction (**q** = 0) [20, 25], whereas, such a rigorous condition would be naturally met for plasmonic polarons. Figure 4c shows a schematic energy diagram of electronic excitations, where there exists a crossover $q_c$ between the competing collective plasmon and e-h pair excitations. As illustrated in this schematic, when $q > q_c$, the collective oscillation mode would decay quickly into e-h pair excitations; thus, the plasmons could survive only in the long-wavelength range ($0 < q < q_c$) [20]. In the case of semimetallic SrIrO$_3$ films, we could evaluate $q_c$ to be less than 15% of the BZ (see details in the Supplementary materials). In addition, in the formalism of Fan-Migdal approximation's many-body perturbation theory, the electron-plasmon coupling matrix elements also exhibit a 1/|**q**| singularity, which is remarkably analogous to the Fröhlich electron-phonon coupling matrix elements [20, 40]. Thus, only rather small momentum (the long-wavelength limit) is most likely to be transferred to electrons in the electron-plasmon coupling for SrIrO$_3$.

Next, to evaluate the coupling strength, we applied the standard Lang-Firsov scenario to interpret our data quantitatively [8]. In this model, the single-particle spectral function follows a Poisson distribution $I_n/I_0 = g^n/n!$, where $I_n$ represents the nth replica band intensity, $I_0$ is the quasiparticle band intensity, and $g$ denotes the coupling strength for ***q*** = 0 (average number of plasmons around the photohole). After removing the background, we could fit the integrated intensities of the ARPES spectra using Poisson distribution (shown in Fig. 4d). Our fitting gives $g \sim 0.26$, corresponding to a Fröhlich coupling constant $\alpha_c \sim 0.4$, according to which we can then deduce the radius of plasmonic polarons in SrIrO$_3$ of approximately 10 Å, as shown in Fig. 4e (see the Supplementary materials for details). This result indicates that the formation of plasmonic polarons indeed provides an important dissipation channel for the ARPES photocurrent. In addition, we have carefully explored the relative intensity of the higher order replica bands and believe that intrinsic plasmons play a dominant role in the replica satellites observed in our experimental results (see the Supplementary materials for details). Furthermore, for weak coupling $\alpha_c \sim 0.4$, the effective mass of Fröhlich polarons can be approximated as $m^*/m_0 = 1/(1-\alpha_c/6)$ [47]. Therefore, the mass renormalization factor from the above relation for three-dimensional Fröhlich polarons is approximately 1.07, which is smaller than the value ~1.23 obtained from the comparisons of the experimental data to single-particle calculations. This behavior can tentatively be attributed to the effect of electron-electron interactions, which is not fully included in the analysis of coupling strength in the polaronic regime [10].

## 4. Conclusion

In common plasmonic metals, plasmons have characteristic energy of approximately 10 eV, in which electron-plasmon scattering is suppressed by the energy-conservation selection [20]. In this regard, semimetallic perovskite SrIrO$_3$ films provide a unique platform where electron-plasmon coupling is kinematically allowed. An exceptional combination of several effects enables us to observe the unique plasmonic polaron. On one hand, plasmons with the suitable energy scale in this semimetal can be well investigated by means of high momentum resolution ARPES. Moreover, the phase space for electron-phonon scattering and e-h pair generation at low binding energies is rather small, which would not lead to prominent spectral broadening, as illustrated in Fig. 4b. On the other hand, the presence of band gap in the band manifold of $SrIrO_3$ effectively avoids the mixup between plasmonic replicas and quasiparticle peaks.

Taking into consideration the energy scale of plasmon in $SrIrO_3$, which is comparable to that of phonon/magnon in this compound, the plasmon is expected to pervasively influence the low-lying electronic structure. Thus, the electron-plasmon decay process should be considered on an equal footing with the acknowledged electron-electron repulsion and other couplings between collective modes and electrons to obtain a complete understanding of the quasiparticle dynamics in semimetallic iridates. Our findings may provide new pathways toward plasmon-assisted bandgap tuning and the manipulation of plasmon polaritons, with potential implications for photonics and plasmonics.

**Conflict of interest**

The authors declare that they have no conflict of interest.

**Acknowledgements**

The authors are grateful to Prof. Donglai Feng for helpful discussions. This work was supported by the National Key R&D Program of the MOST of China (2016YFA0300204) and the National Natural Science Foundation of China (11574337, 11874199, and 11874263). P. W. was supported by the National Basic Research Program of China (2015CB654901). Part of this research used Beamline 03U of the Shanghai Synchron Radiation Facility, which is supported by ME2 project (11227902) from the National Natural Science Foundation of China. D.W.S. is also supported by "Award for Outstanding Member in Youth Innovation Promotion Association CAS."

**Author contributions**

Dawei Shen conceived this research project. Wanling Liu and Zhengtai Liu grew the thin film samples. Songhua Cai and Peng Wang performed the TEM measurements. Zhengtai Liu performed the ARPES measurements. Ruixing Zhou and Gang Li performed the first-principles calculations and analyzed the data. Yekai Song, Zhen Wang and Yi Zheng contributed to the measurement of transport. Zhengtai Liu, Gang Li and Dawei Shen wrote the manuscript with input from all authors.

**Appendix A. Supplementary materials**

Supplementary materials to this article can be found bellow.

**Reference**

[1] Mahan GD. Many-particle physics. Plenum Publishers, New York, 3rd ed, (2000).
[2] Damascelli A, Hussain Z, Shen Z-X. Angle-resolved photoemission studies of the cuprate superconductors. Rev Mod Phys 2003;75:473–541.

[3] Lanzara A, Bogdanov PV, Zhou XJ, et al. Evidence for ubiquitous strong electron-phonon coupling in high-temperature superconductors. Nature 2001;412:510–514.

[4] Shen Z-X, Lanzara A, Ishihara S, et al. Role of the electron-phonon interaction in the strongly correlated cuprate superconductors. Philos Mag B 2002;82:1349–1368.

[5] Yoshida T, Tanaka K, Yagi H, et al. Direct observation of the mass renormalization in $SrVO_3$ by angle resolved photoemission spectroscopy. Phys Rev Lett 2005;95:146404.

[6] Sun Z, Chuang YD, Fedorov AV, et al. Quasiparticlelike peaks, kinks, and electron-phonon coupling at the ($\pi$, 0) regions in the CMR oxide $La_{2-2x}Sr_{1+2x}Mn_2O_7$. Phys Rev Lett 2006;97:056401.

[7] Ye HG, Su ZC, Tang F, et al. Extinction of the zero-phonon line and the first-order phonon sideband in excitonic luminescence of ZnO at room temperature: the self-absorption effect. Sci Bull 2017;62:1525–1529.

[8] Chen, CY, Avila J, Wang, SP, et al. Emergence of interfacial polarons from electron-phonon coupling in graphene/h-BN van der Waals heterostructures. Nano Lett 2018;18:1082–1087.

[9] Lu GM, Zhang RH, Geng, S, et al. Myeloid cell-derived inducible nitric oxide synthase suppresses M1 macrophage polarization. Nat Commun 2015;6:6676.

[10] Wang Z, Walker SM, Tamai A, et al. Tailoring the nature and strength of electron-phonon interactions in the $SrTiO_3$(001) 2D electron liquid. Nat Mater 2016;15:835–839.

[11] Moser S, Moreschini L, Jacimovic J, et al. Tunable polaronic conduction in anatase $TiO_2$. Phys Rev Lett 2013;110:196403.

[12] Hofmann A, Cui XY, Schafer J, et al. Renormalization of bulk magnetic electron states at high binding energies. Nat Phys 2009;102:187204.

[13] Lundqvist BI, Single-particle spectrum of the degenerate electron gas. Phys Kondens Mater 1967;6:206–217.

[14] Ryzhii V. Terahertz plasma waves in gated graphene heterostructures. Jpn J Appl Phys 2006;45:L923–L925.

[15] Rana F. Graphene terahertz plasmon oscillators. IEEE T Nanotechnol 2008;7:91-99.

[16] Liang Y, Yang L. Carrier plasmon induced nonlinear band gap renormalization in two-dimensional semiconductors. Phys Rev Lett 2015;114:063001.

[17] Kheifets AS, Sashin VA, Vos M, et al. Spectral properties of quasiparticles in silicon: A test of many-body theory. Phys Rev B 2003;68:233205.

[18] Caruso F, Lambert H, Giustino F. Band structures of plasmonic polarons. Phys Rev Lett 2015;114:146404.

[19] Caruso F, Giustino F. Spectral fingerprints of electron-plasmon coupling. Phys Rev B 2015;92:045123.

[20] Caruso F, Giustino F. Theory of electron-plasmon coupling in semiconductors. Phys Rev B 2016;94:115208.

[21] Aryasetiawan F, Hedin L, Karlsson K. Multiple plasmon satellites in Na and Al spectral functions from *ab initio* cumulant expansion. Phys Rev Lett

1996;77:2268–2271.

[22] Guzzo M, Lani G, Sottile F, et al. Valence electron photoemission spectrum of semiconductors: *ab initio* description of multiple satellites. Phys Rev Lett 2011;107:166401.

[23] Bostwick A, Speck F, Seyller T, et al. Observation of plasmarons in quasi-freestanding doped graphene. Science 2010;328:999–1002.

[24] Polini M, Asgari R, Borghi G, et al. Plasmons and the spectral function of graphene. Phys Rev B 2008;77:081411(R).

[25] Fröhlich H. Electrons in lattice fields. Adv Phys 1954;3:325–361.

[26] Lee JJ, Schmitt FT, Moore RG, et al. Interfacial mode coupling as the origin of the enhancement of $T_c$ in FeSe films on $SrTiO_3$. Nature 2014;515:245–248.

[27] Giannozzi P, Baroni S, Bonini N, et al. QUANTUM ESPRESSO: a modular and open-source software project for quantum simulations of materials. J Phys-Condens Mat 2009;21:395502.

[28] Noffsinger J, Giustino F, Malone BD, et al. EPW: A program for calculating the electron-phonon coupling using maximally localized Wannier functions. Comput Phys Commun 2010;181:2140–2148.

[29] Pizzi G, Vitale V, Arita R, et al. Wannier90 as a community code: new features and applications. J Phys-Condens Mat 2020;32:165902.

[30] Blaha P, Schwarz K, Madsen GKH, et al. Wien2k, An augmented plane wave plus local orbitals program for calculating crystal properties. Techn. Universitat, Vienna (2019).

[31] Haule K, Yee C-H, Kim K. Dynamical mean-field theory within the full-potential methods: Electronic structure of $CeIrIn_5$, $CeCoIn_5$, and $CeRhIn_5$. Phys Rev B 2010;81:195107.

[32] Ohgushi K, Gotou H, Yagi T, et al. Metal-insulator transition in $Ca_{1-x}Na_xIrO_3$ with post-perovskite structure. Phys Rev B 2006;74:241104(R).

[33] Cao G, Durairaj V, Chikara, S, et al. Non-Fermi-liquid behavior in nearly ferromagnetic $SrIrO_3$ single crystals. Phys Rev B 2007;76:100402(R).

[34] Liu ZT, Li MY, Li QF, et al. Direct observation of the Dirac nodes lifting in semimetallic perovskite $SrIrO_3$ thin films. Sci Rep 2016;6:30309.

[35] Nie YF, King PDC, Kim CH, et al. Interplay of spin-orbit interactions, dimensionality, and octahedral rotations in semimetallic $SrIrO_3$. Phys Rev Lett 2015;114:016401.

[36] Chiang T-C. Photoemission studies of quantum well states in thin films. Surf Sci Rep 2000;39:181–235.

[37] Milun M, Pervan P, Woodruff DP. Quantum well structures in thin metal films: simple model physics in reality? Prog Phys 2002;65:99.

[38] Yoshimatsu K, Horiba K, Kumigashira H, et al. Metallic quantum well states in artificial structures of strongly correlated oxide. Science 2011;333:319–322.

[39] Richard P, Sato T, Souma S, et al. Observation of momentum space semi-localization in Si-doped *β*-$Ga_2O_3$. Appl Phys Lett 2012;101:232105.

[40] Caruso F, Verdi C, Poncé S, et al. Electron-plasmon and electron-phonon satellites in the angle-resolved photoelectron spectra of *n*-doped anatase $TiO_2$. Phys

Rev B 2018;97:165113.

[41] Li YH, Qi RS, Shi RC, et al. Manipulation of surface phonon polaritons in SiC nanorods. Sci Bull 2020;65:820–826.

[42] Rademaker L, Wang Y, Berlijn T, et al. Enhanced superconductivity due to forward scattering in FeSe thin films on $SrTiO_3$ substrates. New J Phys 2016;18:022001.

[43] Ishizaka K, Eguchi R, Tsuda S, et al. Temperature-dependent localized excitations of doped carriers in superconducting diamond. Phys Rev Lett 2008;100:166402.

[44] Lischner J, Palsson GK, Vigil-Fowler D, et al. Satellite band structure in silicon caused by electron-plasmon coupling. Phys Rev B 2015;91:205113.

[45] Riley JM, Caruso F, Verdi C, et al. Crossover from lattice to plasmonic polarons of a spin-polarised electron gas in ferromagnetic EuO. Nat Commun 2018;9:2305.

[46] Pines D, Bohm D. A collective description of electron interactions: II. collective vs individual particle aspects of the interactions. Phys Rev 1952;85:338–353.

[47] Mishchenko AS, Prokof'ev NV, Sakamoto A, et al. Diagrammatic quantum Monte Carlo study of the Fröhlich polaron. Phys Rev B 2000;62:6317.

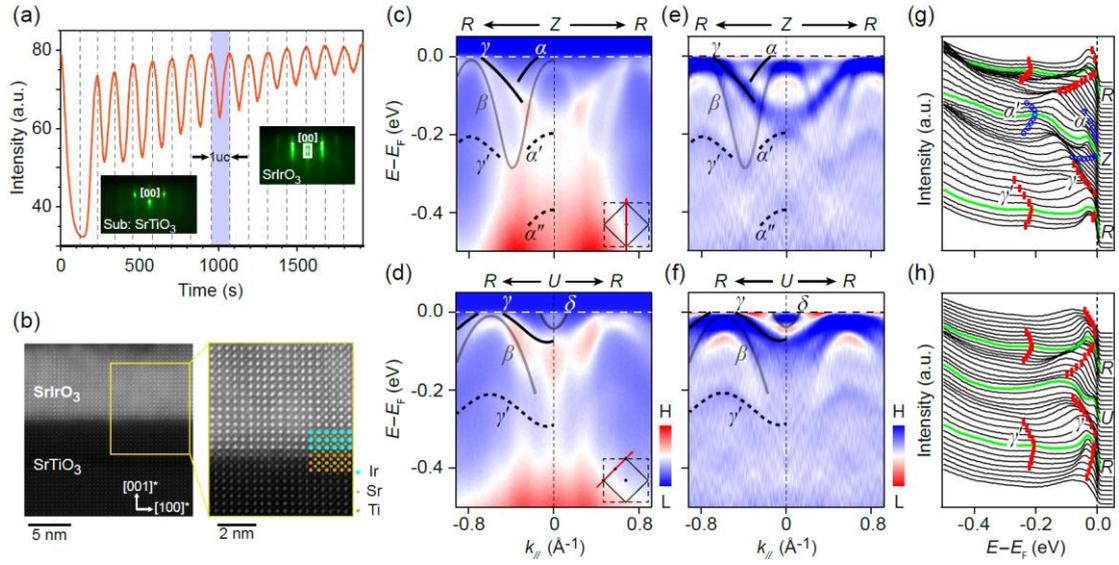

Fig. 1. (Color online) Replica bands of perovskite SrIrO$_3$ films along high-symmetry directions. (a) The RHEED intensity oscillation curve: time dependence of the [00] diffraction spot intensity taken along [100] azimuth direction. Insets demonstrate RHEED patterns of the SrTiO$_3$ substrate prior to growth and SrIrO$_3$ films immediately after growth. (b) (Left) HAADF STEM cross-section z-contrast image of the SrIrO$_3$ films grown on the SrTiO$_3$ substrate. (Right) Close-up view of the yellow box area marked in the left panel. (c and d) Band dispersions along Z–R and U–R high-symmetry directions, respectively, taken at a photon energy of 21.2 eV. For better comparison, both photoemission spectra have been symmetrized with respect to Z and U. Insets in both figures indicate ARPES cuts in the projected two-dimensional BZ, and solid and dashed lines highlight main bands and their high-order replications, respectively. (e–h) Corresponding second derivative images with respect to the energy and EDCs of (c) and (d), respectively.

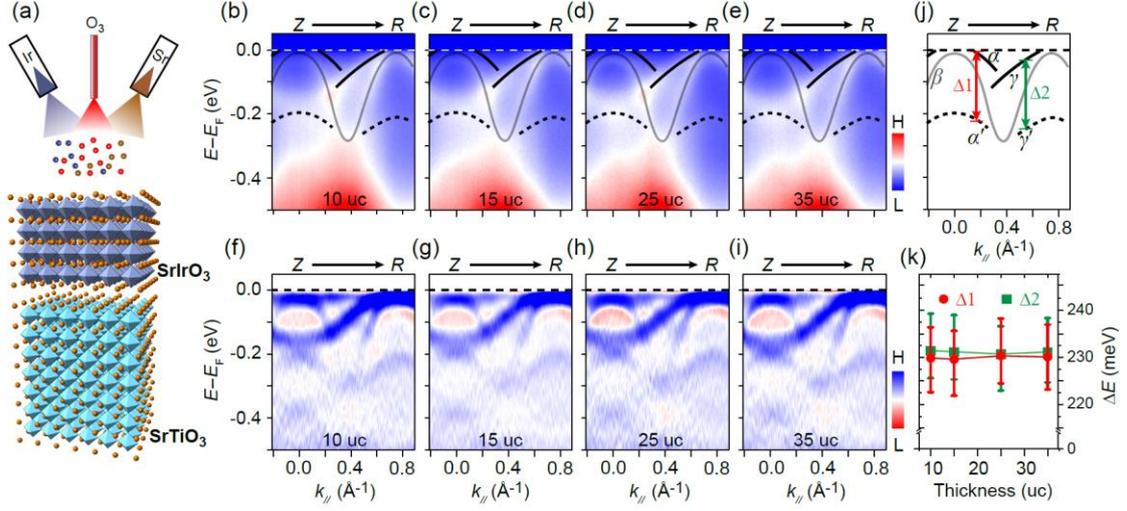

Fig. 2. (Color online) Thickness dependence of main and replica bands in SrIrO$_3$ films. (a) Schematic of a layer-by-layer view of perovskite-type SrIrO$_3$ grown on the SrTiO$_3$ substrate. (b–e) Band dispersions along Z–R of films with different thickness. Solid and dashed lines represent main bands and their replicas, respectively. (f–i) Corresponding second derivatives of (b–e) with respect to the energy. (j) Dispersions of main bands and their first order replicas extracted from the data in (b–e). (k) Thickness dependence of energy shifts between the main and first replica bands for α/α′ (Δ1) and γ/γ′ (Δ2). Error bars are estimated according to the uncertainty of band width due to their spectral broadening in energy.

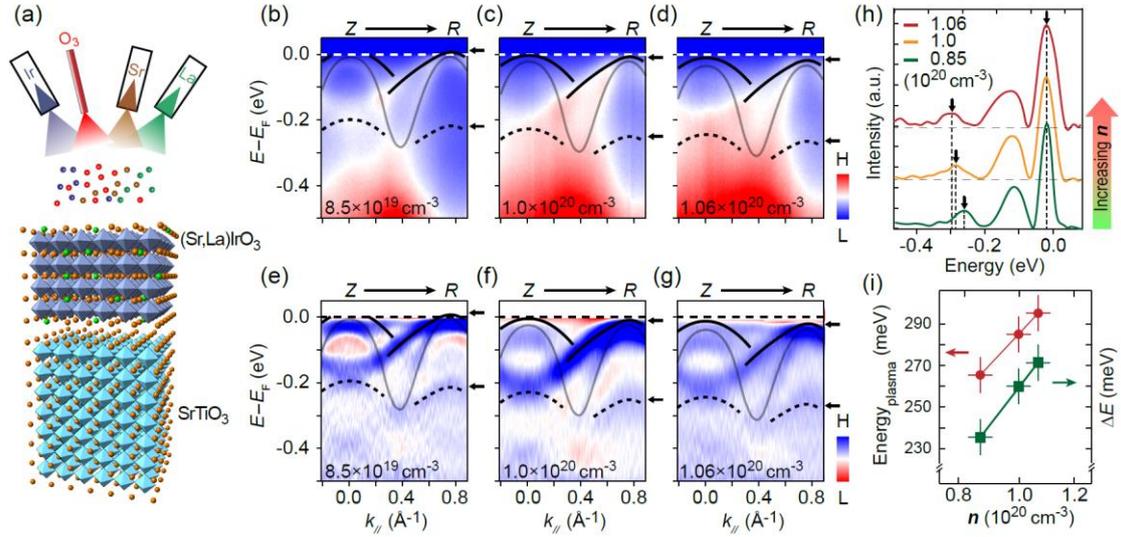

Fig. 3. (Color online) Carrier concentration dependence of main and replica bands in SrIrO$_3$ films. (a) Schematic of fine carrier concentration tuning in La-doped SrIrO$_3$ thin films grown on the SrTiO$_3$ substrate. (b–d) Band dispersions along $Z$–$R$ with different La doping levels. Solid and dashed lines represent main bands and their replicas, respectively. (e–g) Corresponding second derivatives of (b–d) with respect to the energy. (h) Direct comparison among angle-integrated EDCs around $k_{//}$ = 0.56 Å$^{-1}$ for films with different La-doped levels after background subtraction. (Details of background subtraction can be found in the Supplementary materials). Note that, for clarify, we have intentionally aligned chemical potentials of all films. (i) A comparison of the energy separation of the main band and its first replica and plasmon frequency along with the change of carrier concentration. The energy separation and calculated plasmon frequency are labeled with red circles and deep green squares, respectively.

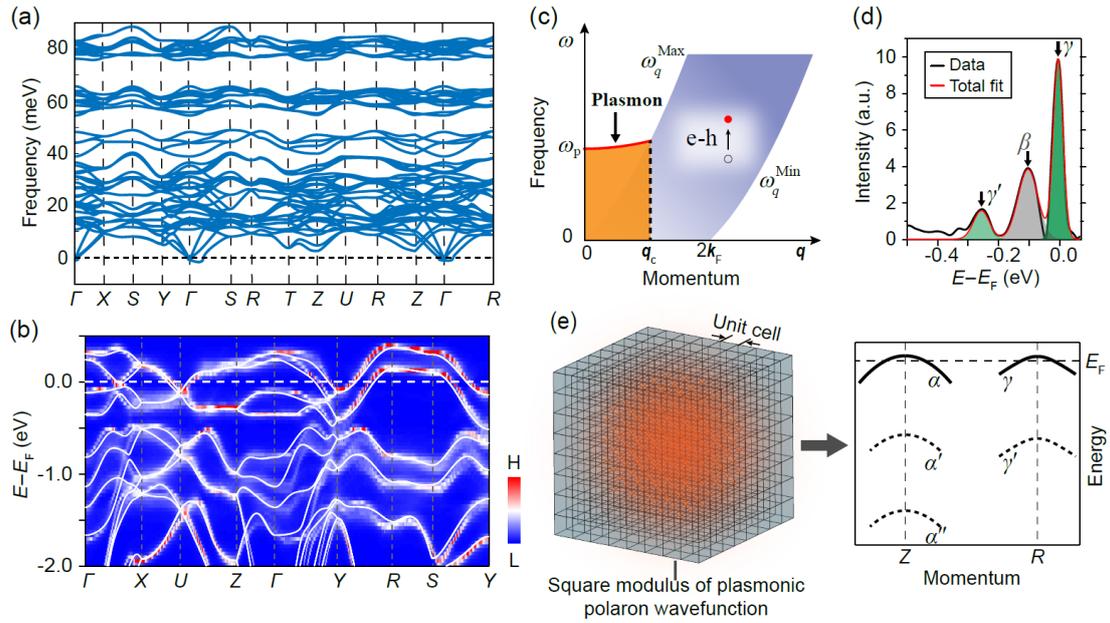

Fig. 4. (Color online) The intrinsic electron-plasmon coupling origin of replica bands. (a) Calculated phonon spectrum along high-symmetry paths in epitaxial $SrIrO_3$ films. (b) Intensity plots of the calculated $k$-resolved spectral function combined with electron-phonon coupling. The band structure (white solid lines) has been superimposed for comparison. (c) Schematic energy diagram of electronic excitations, in which there exists a crossover $q_c$ between the competing collective plasmon and e-h pair excitations. (d) Fitting to the integrated EDCs data around $k_{//} = 0.56$ Å$^{-1}$ with the spline background subtracted (see the Supplementary materials for details). Here, the quasiparticle peak $\gamma$ and plasmon induced replication $\gamma'$ are deep and light green, respectively. The red line shows the full fit. (e) Schematic representation of the square modulus of the plasmonic-polaron wavefunction and band replication induced by electron-plasmon interaction. The minimal cubic structure represents one uc in the crystal lattice.

# Supplementary materials for Electron-plasmon interaction induced plasmonic-polaron band replication in epitaxial perovskite SrIrO$_3$ films

Zhengtai Liu [a,b], Wanling Liu [a,b,c], Ruixiang Zhou [c], Songhua Cai [d,e], Yekai Song [a,b,c], Qi Yao [c], Xiangle Lu [a,b], Jishan Liu [a,b], Zhonghao Liu [a,b], Zhen Wang [e,f], Yi Zheng [e,f], Peng Wang [d,e], Zhi Liu [a,c], Gang Li [c,*], Dawei Shen [a,b,*]

[a] State Key Laboratory of Functional Materials for Informatics, Shanghai Institute of Microsystem and Information Technology (SIMIT), Chinese Academy of Sciences, Shanghai 200050, China
[b] Center of Materials Science and Optoelectronics Engineering, University of Chinese Academy of Sciences, Beijing 100049, China
[c] Division of Photon Science and Condensed Matter Physics, School of Physical Science and Technology, ShanghaiTech University, Shanghai 200031, China
[d] Laboratory of Solid State Microstructures, College of Engineering and Applied Sciences, Nanjing University, Nanjing 210093, China
[e] Collaborative Innovation Center of Advanced Microstructures, Nanjing University, Nanjing 210093, China
[f] Department of Physics, Zhejiang University, Hangzhou 310027, China

*Corresponding authors
Email addresses: ligang@shanghaitech.edu.cn (Gang Li), dwshen@mail.sim.ac.cn (Dawei Shen)

---

## 1. Growth and characterizations of thin films

Thin films of pseudocubic (001) SrIrO$_3$ were synthesized on (001) SrTiO$_3$ substrates by reactive oxide molecular-beam epitaxy (MBE). The growth process was monitored in real-time using *in-situ* reflection high energy electron diffraction (RHEED).

The crystal structure of SrIrO$_3$ films has been *ex-situ* examined by X-ray diffraction (XRD) using a high-resolution Bruke D8 discover diffractometer. Wide-range X-ray $\theta-2\theta$ scans show only SrIrO$_3$ and substrate peaks exist under stoichiometric growth conditions, as shown in Fig. S1a. Laue thickness fringes are present, indicating the high structural quality and sharp interface. The fitted out-of-plane lattice constant of films is 4.0 Å, close to the expected value, suggesting that the films be fully strained (-1.4% lattice mismatch). Furthermore, the best Laue fit to the $\theta-2\theta$ scan around the 001 diffraction peak (Fig. S1b) gives a thickness of 20 uc, which is in an excellent agreement with that determined by counting RHEED oscillations. The rocking curve of the 002 peak of SrIrO$_3$ exhibits a full width at half maxima (FWHM) value which is comparable to the underlying substrate peak (Fig. S1c).

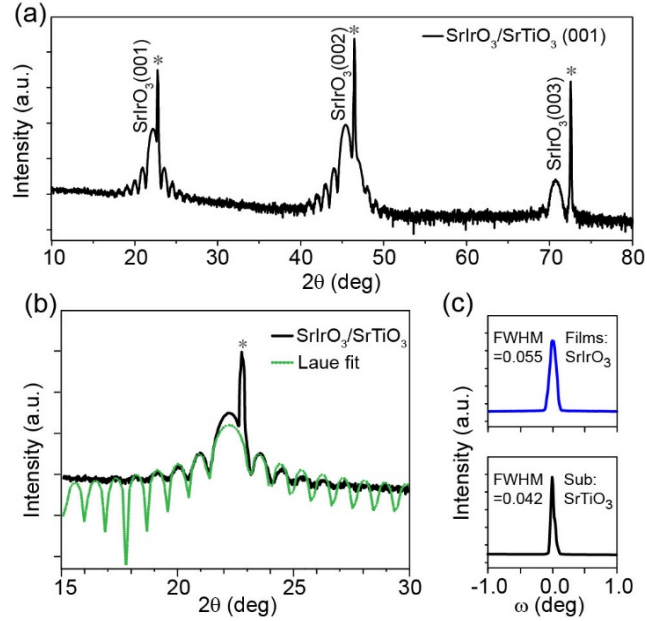

**Fig. S1.** (Color online) X-ray diffraction of SrIrO$_3$ films deposited on SrTiO$_3$ substrate. (a) The typical X-ray diffraction $\theta-2\theta$ scan of SrIrO$_3$ films. (b) The close-up view of the (001) diffraction peak. The green dashed line is the simulation curve. (c), High-resolution X-ray diffraction ω-rocking curves of SrIrO$_3$ films and SrTiO$_3$ substrate.

## 2. Density functional theory calculations

### 2.1. Crystal structure and phonon spectrum

The bulk SrIrO$_3$ studied in this work crystallizes in orthorhombic SG (No. 62). The initial lattice constants, the Wyckoff orbits and the atomic coordinates were taken from a relaxed structure [1] based on the experimental values extracted from powder X-ray diffraction [2]. The density-functional calculations presented in this work was mainly done with quantum espresso package [3]. The kinetic energy cutoff of 680 eV for wave functions was used. The norm-conserving pseudopotential from PSLibrary 0.3.1 [4] and the local density approximation without the spin-orbital coupling (SOC) were adopted in the structure relaxation calculations. The structure was fully relaxed with a 4×4×4 momentum sampling until the force on each atom was smaller than 0.01 eV Å$^{-1}$. The obtained crystal structure parameters: a = 5.6235 Å, b = 5.5683 Å, and c = 7.8879 Å. The relaxed structure is similar to the refined experimental data given in Ref. [1] with a slight change on the displacement of Sr and O(1) along *x*-axis.

Figures S2a and b display the relaxed structure and the corresponding Brillouin zone (BZ). With this structure, the ground-state charge density and the phonon spectrum of SrIrO$_3$ were determined by employing the density functional perturbation theory (DFPT) [5, 6] with a momentum grid of 3×2×3. Our results show that the phonon spectrum displays overall positive values along the high-symmetry path in the first BZ with only negligible negative frequency around *Γ*, indicating a dynamical stable phase of the relaxed structure. The maximum phonon frequency is less than 90 meV in our calculations. The electron-phonon coupling, thus, cannot exceed the maximum frequency of the phonon spectrum. As a result, if there is any electron-phonon interaction induced band replica it will be confined within 90 meV below the Fermi level. However, as clearly indicated by our ARPES measurements, the replica bands *α′* and *γ′* appear around 230 meV, an energy scale much larger than any phonon frequency. We, thus, conclude that the experimentally observed band replica cannot be explained by the phonon degrees of freedom.

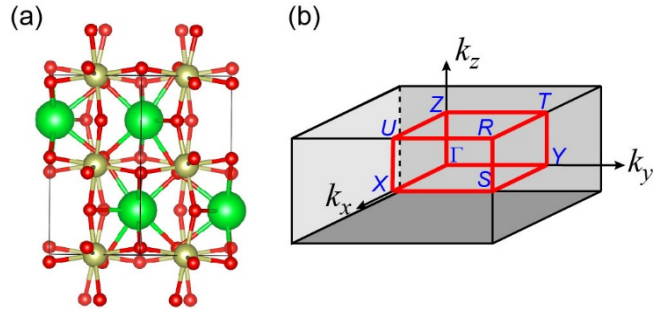

**Fig. S2.** (Color online) (a) The crystal structure and (b) the corresponding BZ of SrIrO$_3$ in space group (No. 62).

Note that, in the phonon study, we have neglected the SOC to accelerate the calculations. As one will see in the next section that the SOC does not play a significant role in the electronic structure, we believe it will likely behave the same in phonon calculations. In particularly, for the quantity we are interested in, *i.e.*, the maximum phonon frequency, we believe the SOC will not lead to a significant change.

*2.2. Electron-phonon interaction*

As electrons can couple to the collective motion of ions, which consequently dresses the motion of electrons with phonon self-energy. As one typical electron-boson coupling, the electron-phonon interaction can result in band replicas separating each with an energy scale defined by that of the bosonic mode. Although it has been excluded, from the previous phonon calculation, the possibility of a phonon origin of the experimentally observed $\alpha'$ and $\gamma'$ bands, it is interesting to see directly the absence of it in the electronic spectrum. And on the other hand, the direct calculation of the phonon-mediated electron spectrum can also help to unveil the possible existence of any collective excitation at even lower energy scale, *i.e.* 0–90 meV.

To determine the electron-phonon coupling in the phonon calculations, the derivative of the self-consistent potential $\partial_{qv}V$ with respect to a collective ionic displacement corresponding to a phonon with branch index v and momentum q needs to be calculated, with which the electron-phonon matrix element: $g_{mn,v}(\mathbf{k}, \mathbf{q}) = (\hbar/2m_0\omega_{\mathbf{q}v})^{1/2}\langle m\mathbf{k} + \mathbf{q}|\partial_{\mathbf{q}v}V|n\mathbf{k}\rangle$, can be straightforwardly obtained. However, SrIrO$_3$ has a relatively large unit-cell. As a consequence, the evaluation of the electron-phonon interaction is numerically very demanding. To this end, we employ the wannier-interpolation scheme and consider only the electronic states of the Ir-*d* orbitals. The phonon frequency is fully taken into account. Thus, the electron-phonon interaction and the resulting electronic self-energy are only relevant to the Ir-*d* states. The methodology description of the corresponding wannierization and interpolation of electron-phonon matrix can be found in Ref. [7]. Our calculations are based on the Electron-Phonon Wannier (EPW) implementation [8].

Figure 4b of the main text displays the electron spectral function of SrIrO$_3$ with the electron-phonon self-energy along different *k*-paths. On top of each plot, the electronic band structure calculated from GGA + SOC is overlaid for comparison. Over the entire BZ we only observed the broadening of the band induced by the electron-phonon interaction. No band replica effect can be clearly visualized. This calculation clearly excludes the phonon-mediated origin of the observed band replica in our ARPES measurement.

*2.3. Electronic structure from DFT*

With the relaxed structure, we now can determine the electronic structure of SrIrO$_3$. To be compatible with the followed correlation study, we here applied the all electron full-potential (linearized) augmented plane-wave plus local orbital (APW+lo) method implemented in Wien2k [9]. The muffin tin radii (RMT) of 2.23, 2.07, 1.69 bohr for Sr, Ir and O were used. The maximum modulus for the reciprocal vectors Kmax was chosen such that RMT·Kmax = 7.0 and a 8×6×8 $k$-mesh in the first BZ was used. The SOC was treated in a second variational way.

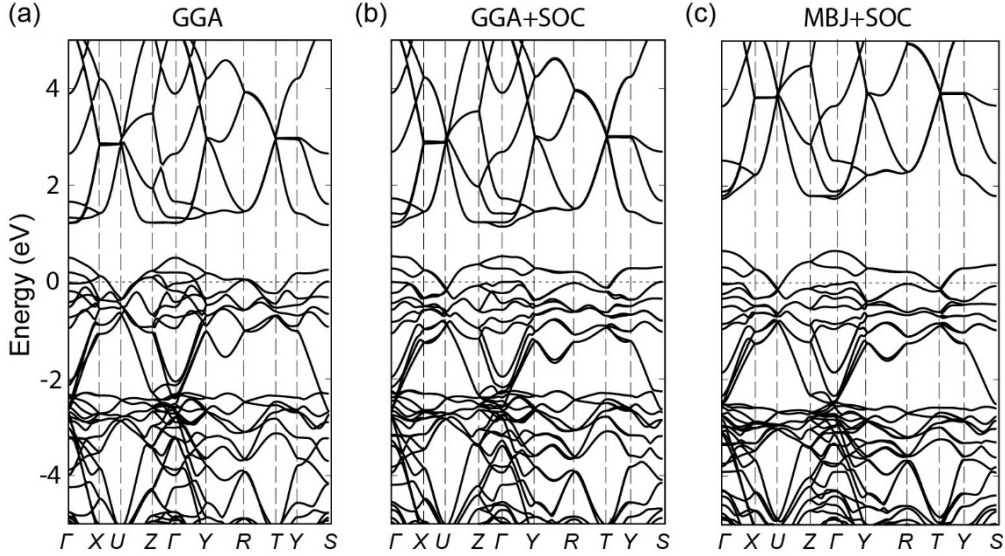

**Fig. S3.** (Color online) The electronic structure of SrIrO$_3$ calculated by (a) GGA, (b) GGA + SOC, and (c) MBJ + SOC.

As displayed in Fig. S3, SrIrO$_3$ is predicted by density-functional theory calculations as a metal with the Ir-$d$ orbitals nearly half occupied. The Ir-$d$ orbitals are separated by an indirect energy gap ~ 0.6 eV in standard GGA calculations, which is nearly unchanged by the further inclusion of SOC. The Fermi level is slightly below the top valence bands. The metallic nature of SrIrO$_3$ is unaffected by the SOC, while a momentum-dependent energy gap is generated around the Fermi level, following which the top two valence bands (the number of bands is actually four when the Kramer's degeneracy is further taken into account) separate from the rest of the valence bands everywhere except for a four-fold band degenerate point close to the $U$-point. As a result, SrIrO$_3$ can be viewed as a Dirac semimetal with a momentum-dependent energy gap induced by the SOC. The four-fold degenerated energy point is a Dirac point protected by the nonsymmorphic symmetries under the presence of Kramer's degeneracy [10]. Any perturbation, including moderate electron-electron interaction, that respects these symmetries cannot lift the band degeneracy.

To see this, we further employed the MBJ exchange potential which provides a better description of the electronic correlations than the standard GGA while still keeps the calculations reasonably fast. It is worthwhile to note that, the correlated $d$-orbits distribute over a large energy window ranging roughly from −4 eV to 4 eV, indicating its delocalized nature. The Ir-$d$ electrons, thus, are not confined with small kinetic energy. As a result, the electron-electron interaction is unlikely to result in any significant change on the electronic structure of this system. As expected, the MBJ result shown in Fig. S3c displays not much difference from the GGA solution shown in Fig. S3b. The energy gap around 1 eV above the Fermi level in Fig. S3b is enlarged and the bands above this gap is further pushed towards higher binding energy. While, the system remains to be a semimetal with the bulk Dirac point around U unaffected by the electron-electron correlations incorporated in the MBJ exchange potential. As stated before, the presence of this Dirac point is symmetry protected. Unless the nonsymmorphic symmetries or the Krameter's degeneracy is destroyed by the electronic correlations and a conventional phase transition is spontaneously triggered, the band degeneracy of the Dirac point will not be lifted.

In addition, as shown in Fig. S3, we have calculated the band dispersion of SrIrO$_3$ in the vicinity of $E_F$ along the $k_z$ direction ($\Gamma$-$Z$). It was found that both $\alpha$ and $\gamma$ bands disperse by no more than 150 meV over the whole Brillouin zone, which is much less than the observed energy shift (~230 meV) of the replica bands. Thus, obviously, states form different $k_z$ values could not account for such band replication phenomenon in SrIrO$_3$.

*2.4. Electron-electron interaction*

To better understand the electronic correlation effect on the electronic structure and to exclude the correlation origin of the replica band, we performed state of the art many-body calculations by means of the dynamical mean field theory (DMFT) with the interaction parameter taken from the cRPA estimation discussed above for the local Ir-$d$ orbitals. The full charge self-consistency with DFT was achieved by employing the embedded-DMFT package [11]. We first obtain the band structure using DFT with the PBE exchange correlation functional in Wien2k. The impurity problem was solved with continuous-time quantum Monte Carlo method [12–15]. DFT+DMFT maps the interacting many-body Hamiltonian (here for Ir-$d$ electrons) defined in non-interacting environment (which represents other bands that are essentially free of interactions) to an interacting impurity problem locally coupled to other non-interacting bands. The local interaction is taken as 3.79 eV as obtained from the cRPA calculations with projection on Ir-$d$ orbitals.

The overall electronic structure, as shown in Fig. S4a, is similar to the DFT results displayed in Figs. S3b and c. To better compare with ARPES, we plotted the electronic spectra along four different $k$-paths around the $R$-point in Figs. S4b–e. The most obvious modification from the electronic correlations appears at the $Z$-point, where the valence top at $Z$-point stays much deeper in energy than that at $R$-point. In DFT calculation, their difference in energy is much smaller, see, for example, the red solid line in Fig. S5e. The electron-electron interactions push the band at $Z$-point more to the high-energy than at $R$-point, resulting in a clear Mexican-hat structure of the top valence band along $R$–$Z$–$R$. In contrast, along $R$–$U$–$R$ the bands seem to be less affected by the electronic correlations, i.e., the correlated spectra shown in Fig. S4e is similar to the DFT band structure. As discussed before, the presence of nonsymmorphic symmetry leads to the emergence of the bulk Dirac point at $k_y = \pi$ plane around $U$-point. The electronic correlation cannot remove these symmetry-protected band degeneracies, although they are dressed by the self-energy now. Similarly, the four-fold band degeneracy at $Z$-point is also preserved in our calculations.

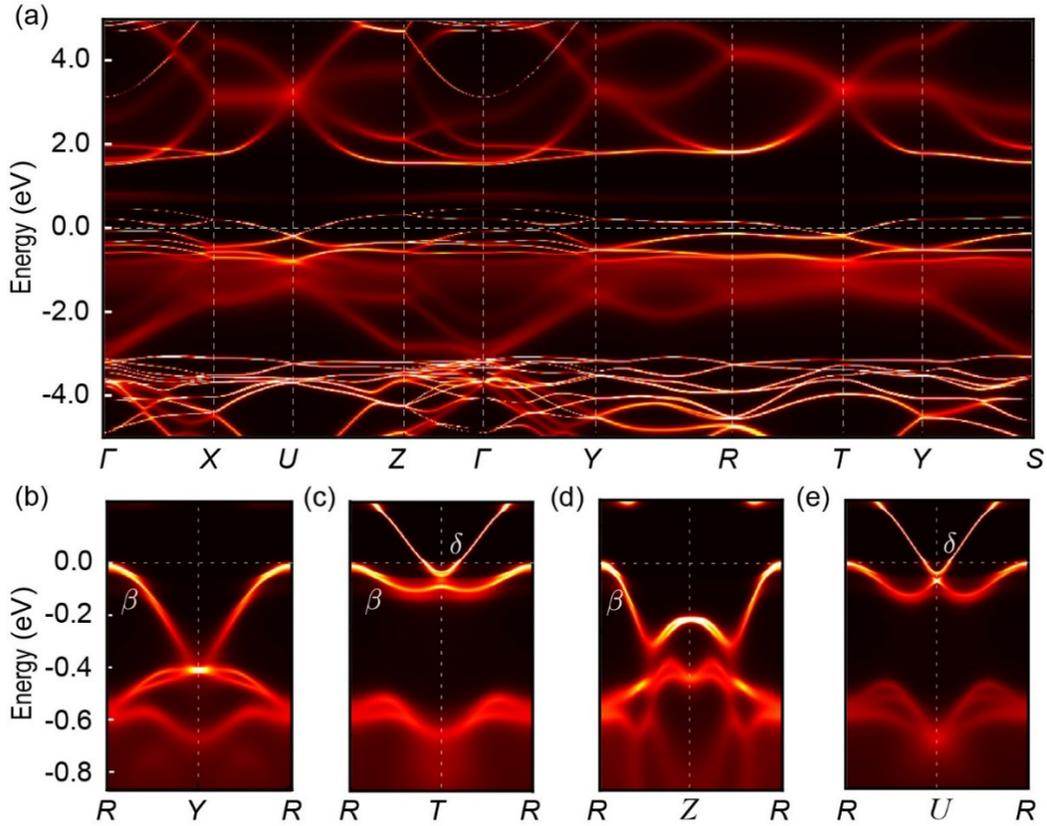

**Fig. S4.** (Color online) (a) The electronic structure of SrIrO$_3$ along the path $\Gamma$–$X$–$U$–$Z$–$\Gamma$–$Y$–$R$–$T$–$Y$–$S$. (b–e) display the zoom-in plots along the similar $k$-path measured in ARPES.

For a better comparison with ARPES spectra, we plot in Figs. S5e–h the corresponding electronic structure from ARPES (the red-blue intensity map), DFT (the solid red line) and DMFT (the red-yellow intensity map). We note that the DFT and the DMFT results are taken for a *k*-path slightly different from that measured in ARPES which are the R–Z–R and R–U–R. They are the parallel *k*-paths of R–Z–R and R–U–R but with a slight reduced $k_z$ value. Away from the $k_y = \pi$ plane the four-fold band degeneracy at Z is lifted. The resulting spectra seem to be better consistent with the ARPES in which two bands (labelled as α and β in the main text) around Z are observed. As clearly seen in Figs. S5a–d, moving away from $k_y = \pi$ plane the two top valence bands along R–Z–R and the first two bands at U-point below the Fermi level are further separated. Note that, in Figs. S5e–h, we have shifted the DFT and DMFT bands to make the bands at R-point consistent with experiment. To achieve this, we shift the Fermi level of the DFT band by −0.1 eV and the DMFT band by −0.18 eV. The DFT bands show an overall better agreement with the ARPES results, in particular along the R–Z–R direction the α and β bands are also in the right energy levels. However, due to the suppression from the electronic correlations, in DMFT the first valence band stays at a much larger energy level at Z-point.

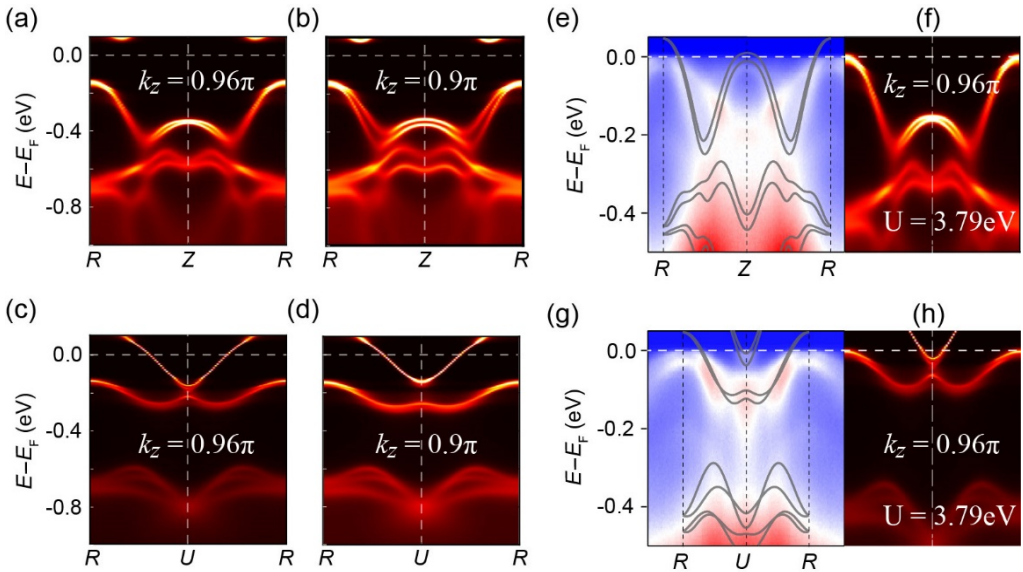

**Fig. S5.** (Color online) (a−d) The left four spectra are taken along the parallel *k*-paths to R−U−R and R−Z−R, but with a different $k_z$ value. The band splitting as a function of $k_z$ can be clearly seen in the comparison. (e−h) The right four plots demonstrate the comparison between the ARPES (red-blue map) with the DFT bands (light grey solid line) and the DMFT spectra (red-yellow map).

In both DFT and DMFT, the γ band is missing. Along the other high-symmetry path, i.e. R–U–R, as the electronic correlations do not play an import role, both DFT and DMFT are consistent with ARPES. The less agreement between DMFT and ARPES is likely due to the improper choice of interaction parameters. Although we adopted the *ab initio* estimation of the U-parameter from cRPA, it was still strongly subject to the approximation of this algorithm. Nevertheless, in both DFT and DMFT we do not observe any signature of band at *R*-point between −0.4 to −0.1 eV, confirming that the replica band γ′ is not of a correlated origin.

## 2.5. Fermi surface volume and carrier concentration

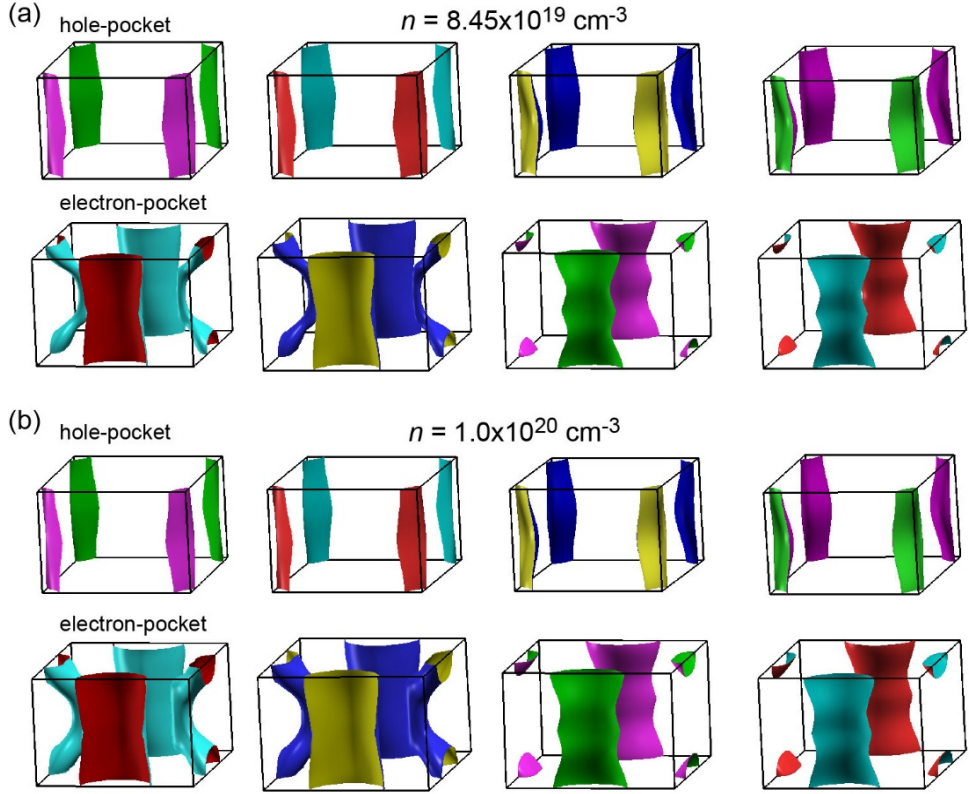

**Fig. S6.** (Color online) The calculated carrier densities for corresponding Fermi surface volumes with different binding energy, respectively. (a) The binding energy ΔE = 24.8 meV, and (b) ΔE = 60.8 meV.

The Fermi surface was calculated over 151×151×151 *k*-mesh in the first BZ, which was found to consist of 4 hole and 4 electron-pockets. They are located at the corner and the edge-center of the BZ, respectively, as shown in Fig. S6. We integrated the Fermi surface volume and estimated the carrier concentration. At the Fermi level, the carrier concentration is $7.2\times10^{19}$ cm$^{-3}$. Slightly increasing the energy level, which corresponded to doping the system with electrons, we obtained the following correspondence between the binding energy and the carrier concentration: ΔE = 24.8 meV, ***n*** = $8.45\times10^{19}$ cm$^{-3}$; ΔE = 60.8 meV, ***n*** = $1.0\times10^{20}$ cm$^{-3}$; ΔE = 92.8 meV, ***n*** = $1.2\times10^{20}$ cm$^{-3}$. Although the detailed electronic structure and the absolute value of the Fermi level strongly depend on the exchange functional adopted in the calculations, our calculation of the Fermi surface volume is highly consistent with experiment estimation. In Fig. S6 we show the Fermi surface at ***n*** = $8.45\times10^{19}$ cm$^{-3}$ and ***n*** = $1.0\times10^{20}$ cm$^{-3}$. With the increase of the binding energy, the carrier concentration increases with the correspondingly hole pockets shrink and the electron pockets expand as shown in Fig. S6. Thus, the contribution of the hole-like carrier density decreases while the electron-like carrier density increases with the doping.

### 3. More details of replica bands along high-symmetry directions

As shown in Fig. S7, there are three hole-like bands $\alpha$, $\beta$ and $\gamma$ around $Z$ and $R$ points, respectively. Figs. S7a–c show the photoemission data taken along $Z-R$ (Cut #1), $Z-U$ (Cut #2) and $U-R$ (Cut #3) high-symmetry directions of the BZ (see Figs. S7g and h). Except for main bands, there also exist two weak hole-like features $\alpha'$ and $\gamma'$ with band tops at the binding energy ~ 200 meV, which surprisingly replicate all features of bands $\alpha$ and $\gamma$, respectively (see Figs. S7a–c and the corresponding EDCs are shown in Figs. S7d–f). Actually, we can even distinguish higher order replica band $\alpha''$ in photoemission intensity images along $Z-R$ and $Z-U$ as shown in Figs. S7a and b (marked by the light blue arrows).

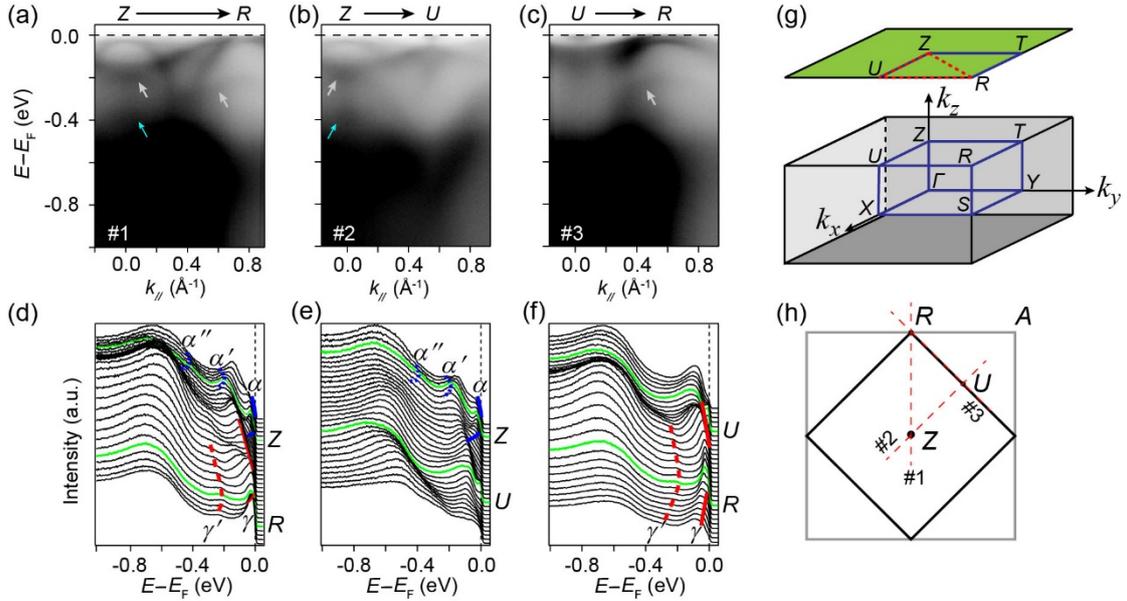

**Fig. S7.** (Color online) Band dispersions along high-symmetry directions. (a–c) The photoemission intensity plots taken along $Z-R$, $Z-U$ and $U-R$ high-symmetry directions, respectively. The arrows indicate the position of replica bands. (d–f) The corresponding EDCs for the photoemission data in (a–c). (g) The three-dimensional BZ and its projection in the $k_x$ and $k_y$ plane. (h) The indication of the cut direction in the projected two-dimensional BZ.

Since replica bands could be observed along all main high-symmetry directions, they are unlikely of photoemission matrix effect origin. Here, the electron-plasmon interaction is dominated by the coupling with the long-wavelength plasmon (q→0), which is of the plasmon frequency $\omega_p = ((ne^2)/(m^*\varepsilon))^{1/2}$. Consequently, all bands, including α, β and γ, should exhibit the same separation to their plasmon satellites. As for β band, its bandwidth (~300 meV) is larger than α and γ bands, exceeding the band-replica separations (~230meV), which probably prevents the formation of band splits.

## 4. The determination of free carrier concentration

The energy of plasmon is proportional to the square root of carrier concentration n in three-dimensional materials, so it is necessary to determine $n$ in an accurate and reliable way to learn the energy of plasmon. We followed a previous literature in which a two-carrier model was successfully applied to determine the carrier concentration in epitaxial SrIrO$_3$ films [16]. In this analysis, the electrical resistivity $\rho$ and Hall coefficient $R_\text{H}$ are given by:

$$\rho = \frac{1}{2ne\mu} \qquad (1)$$

$$R_\text{H} = -\frac{\beta}{ne} \qquad (2)$$

$$\beta = \frac{\mu_e - \mu_h}{\mu_e + \mu_h}, \quad \mu = \frac{\mu_e + \mu_h}{2}, \qquad (3)$$

$$A = (1-\beta^2)\mu^2 \qquad (4)$$

where $n$ is the electron/hole carrier concentration, $e$ is the electron charge, $\mu$ is the average of the electron mobility $\mu_e$ and the hole mobility $\mu_h$, and $\beta$ is their asymmetry. In the low-field limit, the magnetoresistance is roughly proportional to quadratic B$^2$, and $A$ is the corresponding coefficient.

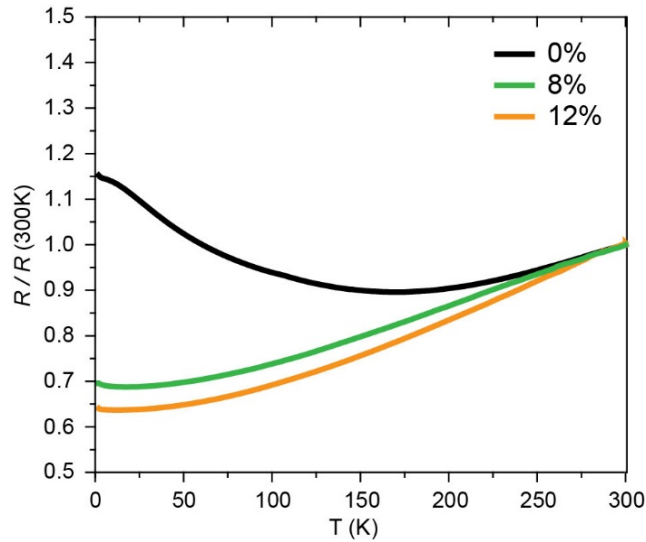

**Fig. S8.** (Color online) Temperature (T) dependence of the normalized resistivity of (Sr,La)IrO$_3$ films.

Figure S8 shows the temperature dependence of normalized resistivity for (Sr,La)IrO$_3$ films with different La doping level. Note that the as-grown SrIrO$_3$ films exhibit typical semimetallic behavior [17, 18], while films with 8% and 12% La doping levels are clearly metallic. We further measured Hall coefficient $R_H$ and magnetoresistance with the magnetic field B normal to the surface at 10 K, as shown in Table I and Fig. S9.

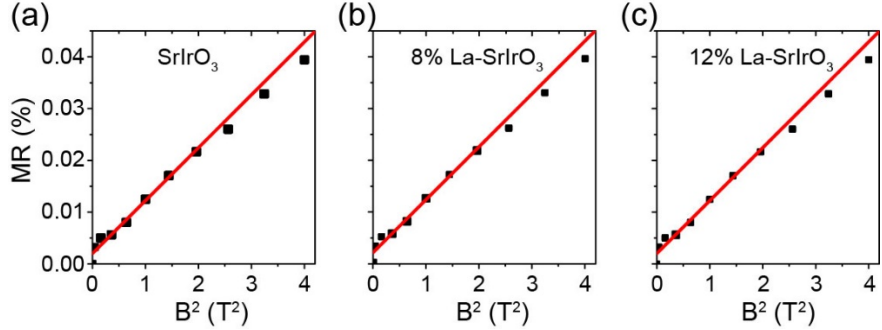

**Fig. S9.** (Color online) Magnetoresistance (black square) versus B$^2$ at 10 K in (a) SrIrO$_3$, (b) 8% and (c) 12% La-doped SrIrO$_3$ films.

Through linear fitting of magnetoresistance in the low field, we have obtained the A value. Finally, according to equations (1–4) above, the carrier concentration at 10 K in the as-grown, 8% and 12% (nominal) La-doping SrIrO$_3$ films could be calculated and then summarized in Table I.

**Table I.** The electrical resistivity, Hall coefficient, the coefficient of magnetoresistance and carrier concentration $n$ (10 K).

| Films | $R$ (μΩ·m) | $-R_H$ (10$^{-3}$cm$^3$/C) | $A$ | $n$ (cm$^{-3}$) |
|---|---|---|---|---|
| SrIrO$_3$ | 3.55×10$^{-6}$ | 15.6 | 1.023×10$^{-4}$ | 8.5×10$^{19}$ |
| 8% La-SrIrO$_3$ | 3.07×10$^{-6}$ | 5.4 | 1.023×10$^{-4}$ | 1.0×10$^{20}$ |
| 12% La-SrIrO$_3$ | 2.90×10$^{-6}$ | 4.0 | 1.023×10$^{-4}$ | 1.06×10$^{20}$ |

In general, the Hall effect in these typical semimetals is usually vanishingly small because the electron and holes contribution cancels. While, in SrIrO$_3$, the Hall coefficient $R_H$ (see Table I) is negative and relatively large, indicative of significant electron-hole asymmetry. In equations (E1)–(E4), the parameter $\beta$ is used to judge the asymmetry of electrons and holes, and the obtained electron-hole asymmetry $\beta$ reaches up to 20%. Moreover, ARPES measurements reveal that the electron-like and hole-like bands in SrIrO$_3$ films have very different effective masses. For example, the $\alpha$ and $\gamma$ hole-like bands bear relatively heavy quasiparticle masses (~ 7 electron masses). In contrast, the electronlike bands are much lighter (~ 0.3 electron masses).

The light electron bands would dominate the transport property, which naturally accounts for the negative sign of the measured Hall coefficient. Therefore, the formation of plasmonic-polarons in SrIrO$_3$ is due to heavy holes (strongly correlated hole bands) dressed by the charge-density fluctuations of electron gas from light electron bands.

## 5. Method of background subtraction

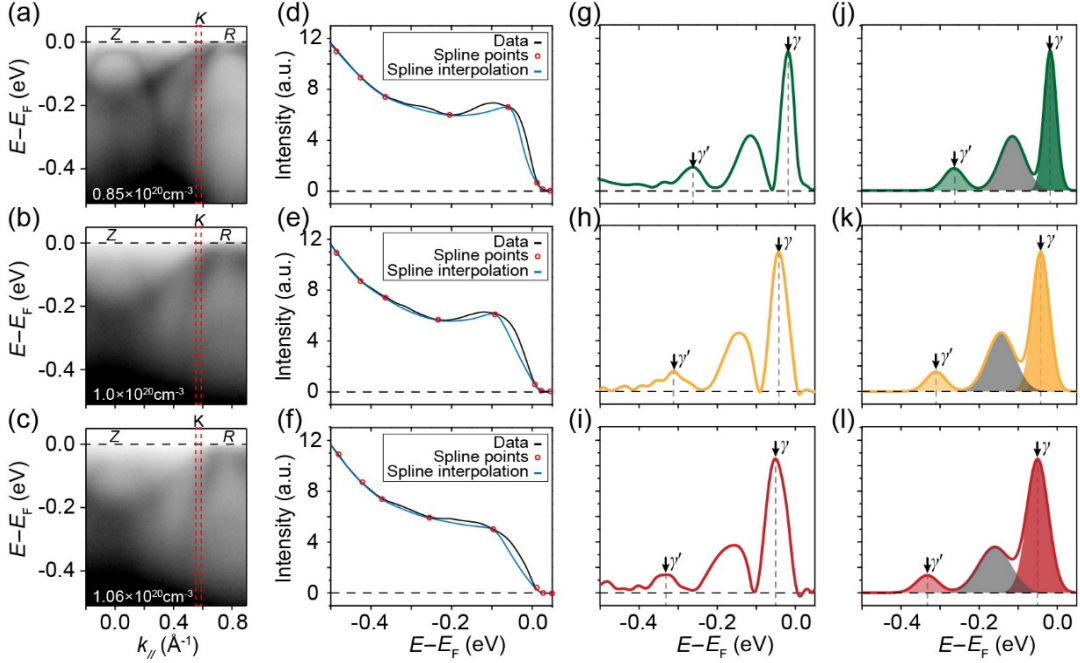

**Fig. S10.** (Color online) Extracting the intensity ratio between the main and replica bands around the $K$ point. (a–c) The photoemission intensity plots taken along $Z-R$ with different carrier concentration. The spectral weight is integrated over the momentum range around $K$ point ($k_{//}$ = 0.56 Å$^{-1}$) indicated by the dotted rectangle to obtain better statistics for a single EDC. (d–f) The integrated EDCs and a method used in the background subtraction for samples with different carrier concentration. Using the red circles as fixed points, we modelled the background using a spline interpolation, plotted with blue curve. (g–i) The integrated EDCs at $K$ point (after background subtraction). Peaks (marked by arrows) corresponding to the main bands and their replicas. Franck-Condon line shape fitting. (j–l) The corresponding fitted EDCs.

In order to estimate the strength of electron-phonon coupling electron-plasmon coupling in materials, we estimate the coherent fraction Z factor from fits to a Franck-Condon model [19–21], which needs to extract the intensity ratio between the main band and its replica. We selected the integrated EDCs around the $K$ point along $Z-R$, as shown in Figs. S10a–c. Because the peaks (main and replica bands) sit on top of a large, non-monotonic background, we chose a spline interpolation background method [following a previous work [22] to extract the intrinsic intensity of main/replica bands, as plotted in Figs. S10d−f. In order to avoid comparatively big discrepancy, we have chosen inflection points near peaks of EDCs as the main spline points, which can eliminate the interference, as far as possible, caused by random selection of spline points.

In addition, we have added extra points to make the background curve subtler and smoother. Here, the red circles indicate the points used to determine the spline interpolation background. Figs. S10g–i show EDCs after subtracting background. Based on that, we fitted the integrated EDCs intensities (Figs. S10j–l) and obtained the Z factor ~ 0.8. The coupling constant is then estimated to be $α_c$ ~ 0.4 from Z factor using the numerical results for the three-dimensional Fröhlich model from Ref. [23].

**6. The determination of intrinsic vs extrinsic plasmons**

It is generally known that satellites in photoemission spectra on the low-kinetic-energy side of main energy bands may be derived from the excitations of plasmons. However, there exist two types of plasmons (intrinsic and extrinsic plasmons) and their intensity in photoelectron spectra competes with each other. In general, the intrinsic excitations of plasmons take place during the ionization event itself, which is the response of the collective oscillation of the conduction electrons to the creation of the positive valence holes by the photoemission. In contrast, the extrinsic plasmons are caused by the Coulomb interaction of conduction electrons propagating through the solid after the photoelectric process. The energy loss of photoelectrons occurs on the movement of the photoelectrons from the site of excitation through the solid to the surface. It is believed that both the extrinsic and intrinsic processes are the same for initial and final state and, thus the photoemission spectra possibly contain both intrinsic and extrinsic plasmon excitations. Here, we have employed a simple method to determine whether the plasmons are intrinsic or extrinsic, which can be made from the relative intensities of successive peaks and is often used in the context of core-level photoemission [24].

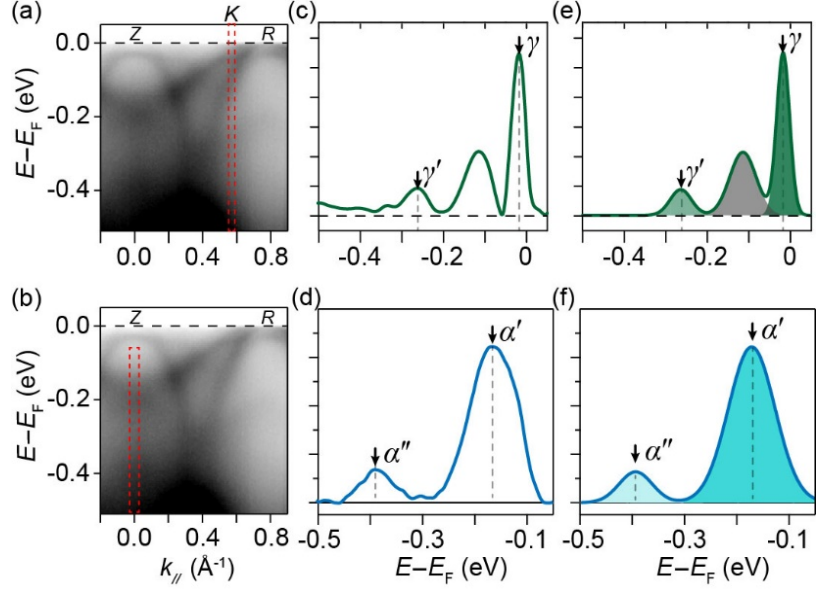

**Fig. S11.** (Color online) Extracting the intensity ratio between $0^{th}$ (main band) and 1st (1st and 2nd) plasmon peaks. (a, b) The spectral weight is integrated over the momentum range around $K$ ($Z$) point indicated by the dotted rectangle. (c, d) The integrated EDCs at $K$ ($Z$) point (after background subtraction). Peaks (marked by arrows) corresponding to the main bands and replicas. (e, f) The corresponding fitted EDCs.

The probability for the creation of the nth bulk plasmon is given by Ref. [25, 26]:

$$I_n/I_0 = b_n + a_n(I_{n-1}/I_0) \quad (5)$$

where $b_n = g^n/n!$ ($g$ denotes the coupling strength) is the creation rate for an intrinsic plasmon, and the second term is the creation rate for an extrinsic plasmon, in which $a_n$ is a slowly varying function of n given by $a_n = a_0(1 + dn)$. In the case of pure extrinsic ($b_n = 0$!) plasmon production the intensity ratios should therefore be nearly constant while for pure intrinsic ($a_n = 0$!) production this ratio should vary as $1/n$.

After removing the background, we have fitted the integrated intensities of ARPES spectra to evaluate the intensity ratio for the bulk plasmons. For clarity, we chose $\gamma$ and $\gamma'$ ($\alpha'$ and $\alpha''$) bands around $K$ ($Z$) point in BZ to extract the intensity ratio between 0th and 1st (1st and 2nd) plasmon peak, which is shown in Figs. S11. Our fitting gives the intensity ratio $\sim$ 0.26 (0.14) for 0th and 1st (1st and 2nd) between $\gamma$ and $\gamma'$ ($\alpha'$ and $\alpha''$) bands. We find the result of intensity ratio is essentially in agreement with the description of pure intrinsic plasmons, in which the ratio $I_2/I_1$ (0.14) is about half the ratio $I_1/I_0$ (0.26). Therefore, we believe that pure intrinsic plasmons play a dominant role in the replica satellites observed in our experimental results, and we are confident that the signatures we discovered are really due to intrinsic plasmonic polaron.

Nevertheless, the ratio $I_2/I_1$ (0.14) slightly higher than half the ratio $I_1/I_0$ (0.26)

potentially implies that there exist extrinsic plasmons, which possess a small proportion of the total.

**7. The determination of critical wave-vector and radius of plasmonic-plasmons**

When larger than the critical wave-vector $q_c$, collective oscillations would decay into electron-hole pair excitations. With a rough approximation $\omega_q = \omega_p$, this value can be given as follows [27]:

$$q_c = \omega_p/v_F \quad (6)$$

The low-energy plasmon frequency with characteristic energies set by the carrier concentration $n$ via $\omega_p = ((ne^2)/(m^*\varepsilon))^{1/2}$, with $m^*$ and $\varepsilon$ being the band effective mass and the high-frequency dielectric constant (4.5 [28, 29]), respectively. Through fitting the electron band dispersion in ARPES data, the Fermi velocity $v_F$ and effective mass could be determined to be ~ 1.2 eV·Å and ~ 0.3 ± 0.05 $m_e$, respectively, in a good agreement with previous reports [17, 18]. The $\omega_p$ is evaluated to be 4.02×10$^{14}$ Hz according to equation (2) in the main text. Thus, the equation (6) can give $q_c$ ~ 0.2 Å$^{-1}$, which is less than 15% of the BZ. Meanwhile, the radius of plasmonic-polarons could be estimated according to Ref. [30], namely, $r_p \cong (3/0.44\alpha)^{1/2}(2m\omega_p/\hbar)^{-1/2}$ ~ 10 Å.

**Reference**

[1] Puggioni D, Rondinelli JM, Comment on "high-pressure synthesis of orthorhombic SrIrO$_3$ perovskite and its positive magnetoresistance". [J Appl Phys 2008;103:103706]. J Appl Phys 2016;119:086102.
[2] Zhao JG, Yang LX, Yu Y, et al. High-pressure synthesis of orthorhombic SrIrO$_3$ perovskite and its positive magnetoresistance. J Appl Phys 2008;103:103706.
[3] Giannozzi P, Baroni S, Bonin N, et al. QUANTUM ESPRESSO: a modular and open-source software project for quantum simulations of materials. J Phys-Condens Mat 2009;21:395502.
[4] PSLibrary. *URL http://theossrv1.epfl.ch/Main/Pseudopotentials*.
[5] Gonze X. Perturbation expansion of variational principles at arbitrary order. Phys Rev A 1995;52:1086–1095.
[6] Gonze X. Adiabatic density-functional perturbation theory. Phys Rev A 1995;52:1096–1114.
[7] Giustino F, Cohen ML, Louie SG. Electron-phonon interaction using wannier

functions. Phys Rev B 2017;76:165108.

[8] Noffsinger J, Giustino F, Malone BD, et al. EPW: A program for calculating the electron-phonon coupling using maximally localized Wannier functions. Comput Phys Commun 2010;181:2140–2148.

[9] Blaha P, Schwarz K, Madsen GKH, et al. Wien2k, an augmented plane wave plus local orbitals program for calculating crystal properties. Techn. Universitat, Vienna (2019).

[10] Chen Y, Kim H-S, Kee H-Y. Topological crystalline semimetals in nonsymmorphic lattices. Phys Rev B 2016;93:155140.

[11] Haule K, Yee C-H, Kim K. Dynamical mean-field theory within the full-potential methods: Electronic structure of $CeIrIn_5$, $CeCoIn_5$, and $CeRhIn_5$. Phys Rev B 2010;81:195107.

[12] Werner P, Comanac A, de Medici L, et al. Continuous-Time solver for quantum impurity models. Phys Rev Lett 2006;97:076405.

[13] Werner P, Millis AJ. Hybridization expansion impurity solver: General formu-lation and application to kondo lattice and two-orbital models. Phys Rev B 2006;74:155107.

[14] Haule K. Quantum Monte Carlo impurity solver for cluster dynamical mean-field theory and electronic structure calculations with adjustable cluster base. Phys Rev B 2007;75:155113.

[15] Gull E, Millis AJ, Lichtenstein, AI, et al. Continuous-time Monte Carlo methods for quantum impurity models. Rev Mod Phys 2011;83:349–404.

[16] Liu J, Chu J-H, Serrao CR, et al. Tuning the electronic properties of $J_{eff} = 1/2$ correlated semimetal in epitaxial perovskite $SrIrO_3$. arXiv 2013;1305:1732.

[17] Liu ZT, Li MY, Li QF, et al. Direct observation of the Dirac nodes lifting in semimetallic perovskite $SrIrO_3$ thin films. Sci Rep 2016;6:30309.

[18] Nie YF, King, PDC, Kim, CH, et al. Interplay of spin-orbit interactions, dimensionality, and octahedral rotations in semimetallic $SrIrO_3$. Phys Rev Lett 2015;114:016401.

[19] Wang Z, Walker SM, Tamai A, et al. Tailoring the nature and strength of electron-phonon interactions in the $SrTiO_3$(001) 2D electron liquid. Nat Mater 2016;15:835–839.

[20] Chen, CY, Avila J, Wang, SP, et al. Emergence of interfacial polarons from electron-phonon coupling in graphene/h-BN van der Waals heterostructures. Nano Lett 2018;18:1082–1087.

[21] Chen CY, Avila J, Frantzeskakis E, et al. Observation of a two-dimensional liquid of Fröhlich polarons at the bare SrTiO$_3$ surface. Nat Commun 2015;6:8585.

[22] Lee JJ, Schmitt FT, Moore RG, et al. Interfacial mode coupling as the origin of the enhancement of $T_c$ in FeSe films on SrTiO$_3$. Nature 2014;515:245–248.

[23] Mishchenko AS, Prokof'ev NV, Sakamoto A, et al. Diagrammatic quantum Monte Carlo study of the Fröhlich polaron. Phys Rev B 2000;62:6317.

[24] Höchst H, Steiner P, Hüfner S. The conduction electron hole coupling in beryllium metal. Phys Lett A 1977;60:69.

[25] Mahan GD. Collective excitations in x-ray spectra of metals. Phys Rev B 1975;11:4814.

[26] Pardee WJ, Mahan GD, Eastman DE, et al. Analysis of surface-plasmon and bulk-plasmon contributions to x-ray photoemission spectra. Phys Rev B 1975;11:3614.

[27] Raether H. Volume plasmons. in: excitation of plasmons and interband transitions by electrons. Springer Tracts in Modern Physics vol.88 (1980).

[28] Seo SSA, Nichols J, Hwang J, et al. Selective growth of epitaxial Sr$_2$IrO$_4$ by controlling plume dimensions in pulsed laser deposition. Appl Phys Lett 2016;109:201901.

[29] Jenderkaa M, Schmidt-Grund R, Grundmann M, et al. Electronic excitations and structure of Li$_2$IrO$_3$ thin films grown on ZrO$_2$:Y (001) substrates. J Appl Phys 2015;117:025304.

[30] Schultz TD. Slow electrons in polar crystals: self-energy, mass, and mobility. Phys Rev 1959;116:526.